\documentclass[epj,final]{svjour}

\usepackage{amssymb}
\usepackage{graphicx}
\usepackage{amsmath}

\begin{document}

\authorrunning{T. Schneider, J. M. Singer}

\titlerunning{Magnetic Field Induced Phase Transitions in ${\bf \mathrm{YBa_2Cu_4O_8}}$}

\title{Magnetic Field Induced Phase Transitions in $\mathsf{YBa_2Cu_4O_8}$}

\author{T. Schneider and J. M. Singer}

\institute{Physikinstitut, Universit\"{a}t Z\"{u}rich, 
Winterthurerstr. 190, CH-8057 Z\"{u}rich, Switzerland}

\mail{\email{jms@physik.unizh.ch}}

\date{Received: \today / Revised version: date}

\abstract{
The $c$-axis resistivity measurements in $\mathrm{YBa_2Cu_4O_8}$
from Hussey et al. for magnetic field
orientations along the $c$-axis
as well as within the $ab$-plane are analyzed and interpreted
using the scaling theory for static and dynamic classical critical phenomena. We
identify a superconductor to normal conductor transition for both field
orientations as well as a normal conductor to insulator  transition at a
critical field $H_c\parallel a$ with dynamical critical exponent $z=1$, leading
to a multicritical point where superconducting, normal conducting and insulating
phases coexist.
}

\PACS{{74.25.Dw}{superconductivity phase diagrams} \and
{74.25.Fy}{superconductivity transport properties}}

\maketitle

Recently it has been demonstrated, that the doping tuned superconductor
to insulator (SI) transition in cu\-pra\-tes can be understood in terms
of quantum critical phenomena in two dimensions \cite{Polonica,Europhys}.
Zero temperature magnetic field driven SI transitions have also been
observed in ultrathin Bi films, and successfully interpreted
in terms of the scaling theory of quantum critical phenomena \cite{Markovic}.
Nevertheless, three important questions concerning the physics of insulating and
superconducting cuprates remain open. One is the nature and dimensionality of
the normal state revealed when superconductivity is supressed by a 
magnetic field \cite{Ando,Malinowski,Levin,Hussey} and the second is the
role of disorder. The third issue concerns the dynamical universality classes
of SI and superconductor to normal state (SN) transitions at finite temperatures.

We address these three issues through an analysis and interpretation of 
recent out-of-plane resistivity measurements $\rho_{c}$ of 
$\mathrm{YBa_2Cu_4O_8}$ in magnetic fields by Hussey et al. \cite{Hussey}, 
using the scaling theory of static and dynamic classical critical phenomena. Since
this material is stoichiometric and, therefore, can be synthesized with
negligible disorder, we consider the pure case. As shown below, the
experimental data for $\rho_{c}(T,{\bf H}\parallel c)$
are consistent with a magnetic field tuned SN transition, while the data for 
$\rho_{c}(T,{\bf H}\parallel a)$ provide strong evidence for a 
multicritical point at the critical field, ${\bf H}_{c}\parallel a,b$, 
where the superconducting, normal conducting and insulating phases coexist. 
Moreover, the existence of the critical field, where the NI transition occurs, 
allows us to determine the dynamical universality class of this transition uniquely:
$z=1$. We attribute the occurrence of the multicritical point to the
comparatively small anisotropy in the correlation lengths, rendering the
critical field for the NI transition to accessible values. For this reason
we expect the multicritical point and the associated NI transition to be
generic for sufficiently clean and homogeneous cuprates with moderate
anisotropy.

The appropriate approach to uncover the phase diagram from conductivity
measurements is the scaling theory of classical dynamic critical phenomena
\cite{Hohenberg}. We now sketch the essential predictions of this theory
in terms of a dimensional analysis.  
A defining characteristic of a superconductor is its broken U(1) or gauge symmetry,
which is reflected in the order parameter $\Psi$. Gauge invariance then implies 
the following identification for the gradient operator
\begin{equation}
i\nabla\Psi \longrightarrow i\nabla\Psi+\frac{2\pi}{\Phi_0}{\bf A}.
\label{EQ1}
\end{equation}
The basic scaling argument, which amounts to a dimensional analysis, states that the
two terms on the right hand side must have the same scaling dimension,
$(\mathrm{Length})^{-1}\equiv L^{-1}$. The dimensionality of the magnetic and
electric field are then expressed as
\begin{equation}
{\bf H} = \nabla\times{\bf A} \propto L^{-2},\quad {\bf E} = \frac{\partial{\bf A}}{\partial t}
\propto (Lt)^{-1}.
\label{EQ2}
\end{equation}
In a superconductor the order parameter $\Psi$ is a complex scalar,
$$\Psi=\mathrm{Re}(\Psi)+i\mathrm{Im}(\Psi),$$
corresponding to a vector with two components. Consequently, the dimensionality
of the order parameter is $n=2$. Based upon the dimensional statement
\begin{equation}
\xi = \xi^\pm_0\vert\epsilon\vert^{-\nu}\propto L,\quad \epsilon=\frac{T-T_c}{T_c},
\label{EQ3}
\end{equation}
where $\pm=\mathrm{sign}(\epsilon)$, we obtain in $D$ dimensions for the free 
energy density the scaling form
\begin{equation}
f = F/(Vk_BT) \propto L^{-D} \propto (\xi^\pm)^{-D},
\label{EQ4}
\end{equation}
and for ${\bf H}\ne 0$
\begin{equation}
f = (\xi^\pm)^{-D}{\cal G(Z)},\quad {\cal Z}=\frac{H(\xi^\pm)^2}{\Phi_0},
\label{EQ5}
\end{equation}
due to Eq. (\ref{EQ2}). 
${\cal G}$ is an universal scaling function of its argument ${\cal Z}$.
An extension to 3D anisotropic materials, such as cuprates,
is straightforward \cite{SchneiderEPJ}:
\begin{equation}
f = \left(\xi^\pm_x\xi^\pm_y\xi^\pm_z\right)^{-1}{\cal G}({\cal Z}),
\end{equation}
where the indices 
$x,y,z$ denote the corresponding crystallographic $b,a,c$-axes of the cuprates, and
\begin{eqnarray}
{\bf H}&=&H(0,\sin\delta,\cos\delta) : \cr &&{\cal Z} = \frac{(\xi^\pm_x)^2}{\Phi_0}
\sqrt{\left(\frac{\xi_z}{\xi_x}\right)^2H_y^2+\left(\frac{\xi_y}{\xi_x}\right)^2H_z^2},\cr
{\bf H}&=&H(\cos\phi,\sin\phi,0) : \cr &&{\cal Z} = \frac{(\xi^\pm_z)^2}{\Phi_0}
\sqrt{\left(\frac{\xi_y}{\xi_z}\right)^2H_x^2+\left(\frac{\xi_x}{\xi_z}\right)^2H_y^2}.
\label{EQ7}
\end{eqnarray}
Using this scaling form of the free energy density magnetization 
\cite{SS_physica,Hubbard} and
magnetic torque data \cite{SchneiderEPJ,Hofer} have been successfully analyzed.

Of particular interest in the present context is the conductivity $\sigma$.
From the dimension of the current,
\begin{equation}
{\bf J} = \frac{\partial f}{\partial{\bf A}} \propto L^{-D+1},
\label{EQ18}
\end{equation}
and the electric field (see Eq. (\ref{EQ2})),
we obtain for the electric conductivity 
\begin{equation}
\sigma = \frac{J}{E} \propto tL^{2-D} \propto \xi^{2-D+z},\quad t\propto\xi^z,
\label{EQ19}
\end{equation}
since the scaling dimension of time is fixed by
$$t\propto L^z\propto\xi^z.$$
\begin{figure}[h]
\centering
\includegraphics[width=7cm]{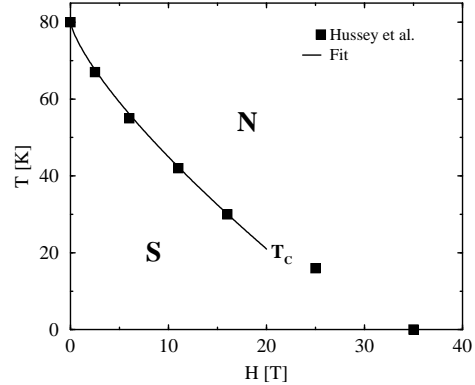}
\caption{$T$-$H$-phase diagram for Y-124, ${\bf H}\parallel c$. The data points
$T_c$ vs. $H_c$ have been deduced from [7]. 
The solid curve corresponds to the limiting behavior given by Eq. (\ref{EQTCBC}).
\label{figtcbc}}
\end{figure}
\noindent The relaxation time $\tau$ describes the rate at which the system relaxes to
equilibrium. $\tau$ diverges at the transition and
the dynamic critical exponent $z$ is defined as
$$\tau\propto\xi_\tau\propto\xi^z\propto\vert\epsilon\vert^{-z\nu}.$$
For $H\ne 0$ the scaling expression for the conductivity reads as
\begin{equation}
\sigma(T,H) = \xi^{2-D+z}{\cal G(Z)},\quad {\cal Z}=\frac{H(\xi^\pm)^2}{\Phi_0}.
\label{EQ20}
\end{equation}
Supposing then that there is a critical point at ${\cal Z}={\cal Z}_c$ with
\begin{equation}
2-D+z = 0,
\label{EQ21}
\end{equation}
the curves $\sigma$ versus $H$ recorded
at different temperatures $T$ will cross at $H=H_c$, where ${\cal Z}={\cal Z}_c$.
The extension to anisotropic systems reads as
\begin{equation}
\sigma_{ii} \propto \frac{\xi_i\xi_\tau}{\xi_j\xi_k}{\cal G(Z)},
\label{EQSIG}
\end{equation}
with $(i,j,k)\equiv(x,y,z)$ and ${\cal Z}$ given by Eq. (\ref{EQ7}).

\begin{figure}[h]
\centering
\includegraphics[width=6cm]{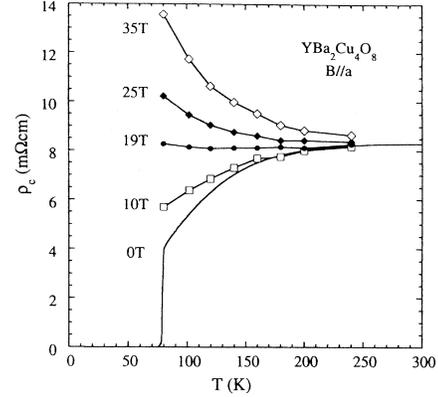}
\caption{$\rho_c(T)$ curves for various fields
${\bf H}\parallel a$, taken from [7].
\label{fighuss1}}
\end{figure}

We are now prepared to analyze the resistivity data of Hussey et al. 
\cite{Hussey}.
At $H=0$, the material is supposed to undergo a continuous SN transition belonging
to the $3D$-$XY$ universality class, and accordingly
\begin{equation}
\xi_{x,y,z} = \xi_{0,x,y,z}\vert\epsilon\vert^{-\nu},
\ \xi_\tau\propto\vert\epsilon\vert^{-z\nu},\ \nu\approx 2/3.
\end{equation}
If the transition occurs in finite fields, then ${\cal Z}={\cal Z}_c$, and
close to $T_c=T_c(H=0)$ the phase transition line is given by
\begin{equation}
H_{c,i}=\frac{{\cal Z}_c\Phi_0}{\xi_{0,j}\xi_{0,k}}\left(\frac{T-T_c}{T_c}\right)^{2\nu},
\ (i,j,k)\equiv(x,y,z).
\label{EQHC2}
\end{equation}
From Fig. \ref{figtcbc} it is seen that the resulting behavior for $H_{c,z}$ (i.e.
${\bf H}\parallel c$), namely
\begin{equation}
T_c(H) = T_c(H=0)\left(1-0.078H^{3/4}\right),
\label{EQTCBC}
\end{equation}
agrees remarkably well with the experimental data. 
Similarly, for $H_{c,y}$ (${\bf H}\parallel a$)
we obtain the estimate (using the  
two points $T_c(H=0)=80K$, $T_c(H=35T)=65K$ measured by \cite{Hussey})
\begin{equation}
T_c(H) \approx T_c(H=0)\left(1-0.013H^{3/4}\right),
\label{EQTCBA}
\end{equation}
which has been included in the phase diagram shown Fig. \ref{figtcba}. 
\begin{figure}[h]
\centering
\includegraphics[width=5.7cm]{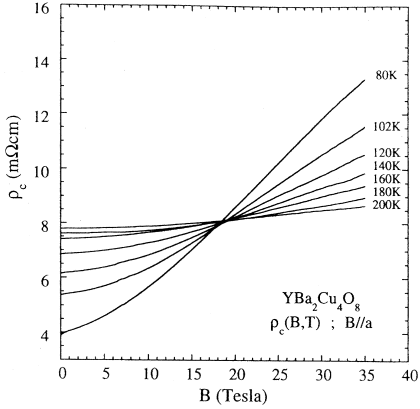}
\caption{$\rho_c$ versus $H$, ${\bf H}\parallel a$ for various temperatures,
taken from [7]. 
\label{fighuss2}}
\end{figure}
Combining Eqs. (\ref{EQHC2}), (\ref{EQTCBC}) and (\ref{EQTCBA}), we obtain for the 
$yz$-anisotropy of the correlation lengths the estimate
\begin{equation}
\frac{\xi_y}{\xi_z}\approx\left(\frac{0.078}{0.013}\right)^{4/3} \approx 11,
\label{EQANISO1}
\end{equation}
which is close to the value obtained from magnetic torque measurements \cite{Zech}.
\begin{figure}[h]
\centering
\includegraphics[width=7cm]{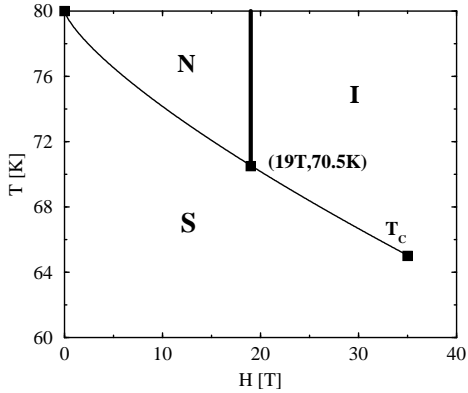}
\caption{Sketch of the $T$-$H$-phase diagram, Y-124, ${\bf H}\parallel a$, 
deduced from the experimental data shown in Fig. \ref{fighuss1} and \ref{fighuss2}.
The solid line corresponds to Eq. (\ref{EQTCBA}). 
S: Superconductor; I: Insulator; N: Normal conductor.
\label{figtcba}}
\end{figure}

\noindent
According to Figs. \ref{fighuss1} and \ref{fighuss2}, showing $\rho_c$ vs. temperature
for various fields $({\bf H}\parallel a)$ as well as $\rho_c$ vs. $H$
$({\bf H}\parallel a)$
for various temperatures between $80K$ and $200K$, the behavior of the
$c$-axis resistivity for a field oriented along the $a$-axis 
differs drastically from the
$({\bf H}\parallel c)$ data. Indeed, it is seen that 
in the normal state $T_c<T<200K$ for $({\bf H}\parallel a)$ a 
NI transition occurs.  In particular,
all $\rho_c({\bf H}\parallel a)$ curves go through a 
single crossing point (Fig. \ref{fighuss2}) at 
$$H_c\approx 19T,\quad \rho_{c,c}\approx 8m\Omega cm,$$
which is the value where $\rho_c(T)$ becomes essentially temperature independent (Fig.
\ref{fighuss1}). Nevertheless, 
there is still a transition into a superconducting regime beyond
$H_c\approx 19T$ for ${\bf H}\parallel a$ (e.g. $T_c(35T)=65K$).
One clearly observes in Fig. \ref{fighuss1} that at low magnetic fields 
($H<H_c\approx19T$), as the
temperature is reduced, $\rho_c(T)$ shows a drop from its normal state value
$\rho_c\approx 8m\Omega cm$. For $H>H_c\approx 19T$ and $T>T_c(H)$, $\rho_c$
raises with decreasing temperature, signalling the onset of insulating behavior.
The resulting phase diagram, showing SN, SI and NI transitions as well as a
multicritical point, where the superconducting, normal conducting and insulating
phases can coexist, is drawn in Fig. \ref{figtcba}. On physical grounds one expects
that the NI transition line will have a critical endpoint in the normal state, too.

The existence of the crossing point (critical field) in the  bulk material 
($D=3$) (Fig. \ref{fighuss2}) implies according to Eq. (\ref{EQ21})  
that the dyamical critical exponent of the NI transition is $z=1$.
As a consequence, the scaling form of the conductivity (Eq. (\ref{EQSIG}))
reduces for ${\bf H}\parallel a,b$ to
\begin{equation}
\sigma_{zz}\propto {\cal G}({\cal Z}_c) \ \mathrm{for} \ z=1, \ D=3,
\end{equation}
where
\begin{equation}
{\cal Z}_c = \frac{\xi_z\xi_yH_c}{\Phi_0}\sqrt{\cos^2(\phi)+\left(\frac{\xi_x}
{\xi_y}\right)^2\sin^2(\phi)}.
\end{equation}
Thus, 
curves $\rho_c$ vs. $H$ (${\bf H}\parallel a$), recorded at different temperatures $T$,
exhibit a crossing point at $H=H_c$, where ${\cal Z}={\cal Z}_c$, in
agreement with the experiment (Fig. \ref{fighuss2}).
\begin{figure}[h]
\centering
\includegraphics[width=7cm]{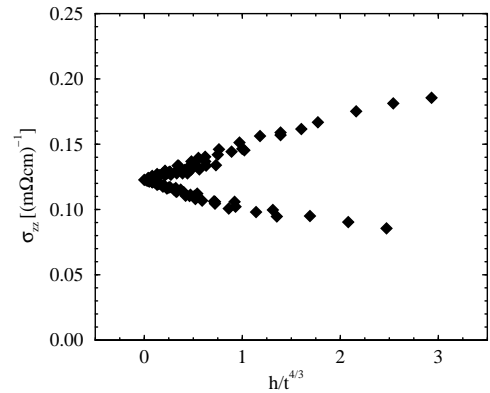}
\caption{$\sigma_{zz}$ vs. $h/t^{4/3}$ for the data shown in Fig. \ref{fighuss2}.
Upper branch: $H<H_c$; lower branch: $H>H_c$.
\label{figkollaps}}
\end{figure}
\noindent
Moreover, if the data for $H\parallel a$, as shown in  Fig. \ref{fighuss2}, 
are replotted according to
\begin{equation}
\sigma_{zz} = {\cal F}\left(h/t^{2\nu}\right),
\label{EQKOLL}
\end{equation}
with $h=\vert H-H_c\vert/H_c$ and $t=\vert T-T_c(H_c)\vert/T_c(H_c)$,
they collapse, as shown in Fig. \ref{figkollaps},
onto two branches (using $T_c(H_c)\approx 70.5K$, 
$H_c\approx19T$ and  $\nu\approx 2/3$).

Finally, close to the NI transition (${\cal Z}={\cal Z}_c$) we obtain
\begin{equation}
\sigma_{zz} \approx \sigma_{zz}({\cal Z}_c)+\frac{\partial\sigma_{zz}}{\partial {\cal Z}}
\bigg\vert_{{\cal Z}={\cal Z}_c} ({\cal Z}-{\cal Z}_c),
\end{equation}
where
\begin{equation}
{\cal Z}-{\cal Z}_c=(H-H_c)\frac{\xi_z\xi_y}{\Phi_0}
\sqrt{\cos^2\phi+\left(\frac{\xi_x}{\xi_y}\right)^2\sin^2\phi}.
\end{equation}
Thus, provided that $\xi_y\ne \xi_x$, both, out-of-plane conductivity and 
resistivity will depend on the angle $\phi$, so that close to the multicritical
point
\begin{eqnarray}
\Delta\rho_{zz} &=& \frac{1}{\sigma_{zz}-\sigma_{zz}({\cal Z}_c)} 
= \bigg(\frac{\partial\sigma_{zz}}{\partial{\cal Z}}\bigg\vert_{{\cal Z}={\cal Z}_c}
(H-H_c)\cr
&&\qquad\qquad \cdot\frac{\xi_z\xi_y}{\Phi_0} \sqrt{\cos^2\phi+\left(\frac{\xi_x}{\xi_y}\right)^2
\sin^2\phi}\bigg)^{-1}.
\label{EQ_RHOZZ}
\end{eqnarray}

\begin{figure}
\centering
\includegraphics[width=7cm]{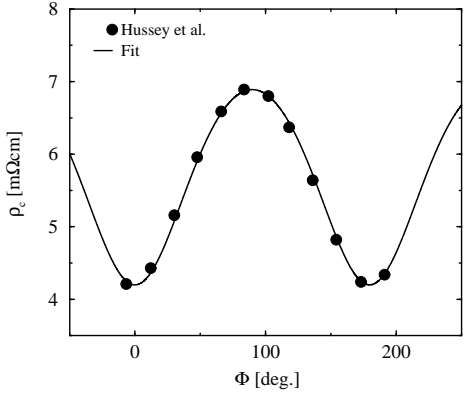}
\caption{Angular dependence of the $c$-axis resistance $\rho_c$ recorded at $T=85K$
and field $H=15T$. The field is rotated in the $ab$-plane.
Experimental data (circles) are taken from [7], the solid line is a fit to 
expression (\ref{EQ_RHOZZ}), yielding $\xi_x/\xi_y=1.27$.
\label{figfit}}
\end{figure}

\noindent
Fig. \ref{figfit} shows a fit of the Y-124 data to Eq. (\ref{EQ_RHOZZ}),
this expression describes the experimental data very well, 
yielding $\xi_x/\xi_y=1.27$. 
Given the estimate $\xi_y/\xi_z\approx 11$
(Eq. (\ref{EQANISO1})) and noting that 
at the crossing point the magnetic field has to scale (Eq. (\ref{EQSIG})) as
\begin{equation}
{H_{c,z}}/{H_{c,y}} = {\xi_y}/{\xi_z},
\end{equation}
it is readily seen that 
with $H_{c,y}\approx 19T$ and ${\bf H}\parallel (z=c)$ a magnetic 
field driven NI transition might occur for rather high fields only. 
Indeed, according to the experimental phase diagram shown in Fig. \ref{figtcbc}, 
there is no such NI transition up to $H=35T$, where the superconducting phase 
dissappears even at $T=0$ and a magnetic field driven 
SN quantum phase transition occurs.

To summarize, we have shown that the magnetic field and temperature dependence
of the $c$-axis resistivity in $\mathrm{YBa_2Cu_4O_8}$, recorded for various
magnetic field orientations, can be understood in terms of the scaling theory
of static and dynamic classical critical phenomena.
For ${\bf H}\parallel c$ we identified a SN transition. 
For ${\bf H}\parallel a$ we provide considerable evidence 
for the occurence of a finite temperature NI
transition in $D=3$ with the dynamical critical exponent $z=1$,
as well as for a multicritical point, where superconducting, normal conducting
and insulating phases can coexist.   The occurence of
this multicritical point must be attributed to 
the comparatively small anisotropy, characterized by the correlation length ratio
$\xi_y/\xi_z$ ($\xi_y/\xi_z\approx11$),
rendering the critical field to accessible values. 
For this reason, the multicritical point and the associated NI 
transition are expected to be generic for sufficiently clean, homogeneous
and almost optimally doped
cuprates with moderate anisotropy $\xi_y/\xi_z$ and $\xi_x/\xi_z$.

\bigskip

We benefitted from discussions with H. Keller, J. Hofer and M. Willemin.
Part of the work was supported by the Swiss National Science Foundation.

\end{document}